\documentclass[aps,twocolumn,prb,superscriptaddress,floatfix,showpacs]{revtex4}
\usepackage{amsmath}
\usepackage{graphicx}
\usepackage{subfigure}
\usepackage{color}
\usepackage{epstopdf}

\def\lsim{\mathrel{\mathpalette\gl@align<}}
\def\gsim{\mathrel{\mathpalette\gl@align>}}
\def\gl@align#1#2{\lower.6ex\vbox
{\baselineskip\z@skip\lineskip\z@
\ialign{$\m@th#1\hfil##\hfil$\crcr#2\crcr\sim\crcr}}}

\makeatother

\newcommand\ba{\begin{eqnarray}}
\newcommand\ea{\end{eqnarray}}
\newcommand\be{\begin{equation}}
\newcommand\ee{\end{equation}}
\newcommand\bi{\bibitem}

\begin{document}

\title{Sudden quenches in quasiperiodic Ising model}

\author{Uma Divakaran}
\affiliation{Department of Physics, Indian Institute of Technology Palakkad, Palakkad, 678557, India}

\date{\today}

\begin{abstract}

We present here the non-equilibrium dynamics of the recently studied quasiperiodic Ising model \cite{anushya17}.
The zero temperature phase diagram of this model mainly consists of three phases, where each of these three phases
can have extended, localized or critically delocalized low energy excited states.
We explore the nature of excitations in these different phases by studying the evolution of entanglement entropy 
after performing quenches of different strengths to different phases.
Our results on non-equilibrium dynamics of entanglement entropy are concurrent with the nature of excitations
discussed in Ref. \onlinecite{anushya17} in each phase.

\end{abstract}


\maketitle


\section{Introduction}

Quasiperiodicity in a Hamiltonian introduces many interesting properties to the states of the system.
One of them being the simultaneous presence of extended as well as localized states even in one dimensions.
It is well known that a disordered system 
can have only localized states in one and two dimensions, as proposed by Anderson\cite{anderson58}. 
The possibility of extended states exist only in three dimensions.
On the other hand, quasiperiodicity in many models can actually bring in a combination of localized and extended states
even in one-dimensions as such lattices have a spatial ordering which is intermediate between a periodic 
ordering and completely random or disordered ordering. 
For example, one dimensional Aubry Andre (AA) model is essentially an XX chain in presence of
quasiperiodic transverse field\cite{aa,harper}. This model can be solved owing to its self-dual nature under suitable transformations
to momentum space. For certain parameter range, all the states of the system are extended
states whereas in the remaining range of parameters, all the states are localized, allowing a possibility of phase transition
dictated by change in the nature of states. Detailed studies on the static properties of Aubry Andre model
have been done earlier\cite{aa,harper,modungo09,deissler11}, with the recent addition of non-equilibrium dynamics giving a better picture of the model.  \cite{roosz14}
As opposed to before, quasiperiodicity now
is not simply a matter of theoretical interest, but also of experimental
relevance owing to the ability of generating such lattices in optical experiments using lasers of incommensurate wavelengths \cite{bloch08,qp_lattice}.
In fact, single particle localization in 
quasiperiodic lattices has been observed experimentally \cite{qp_single}.
Such optical experiments offer a well controlled tool to study the phenomena of localization and symmetry breaking present in condensed matter systems.

Quasiperiodicity can also bring in dynamically stable long-range orders which are otherwise forbidden
in equilibrium.
Existence of dynamically stable long range ordered states was first proposed by Huse $et.al$ \cite{huse13} in the context of Many Body Localization.
This type of localization driven 
protected state phenomena may be responsible for a sharp topological phase even in highly excited state \cite{bahri15}.
With these existing results or proposals, Anushya $et.al$ studied a variant of Aubry Andre model \cite{anushya17}, which is quasiperiodic Ising 
model in a transverse field (QPTIM).
Such a quasiperiodic Hamiltonian also has a combination of localized and extended states for certain Hamiltonian parameters and demonstrated the
existence of dynamically stable long range orders which are not present in equilibrium. We shall briefly discuss this model in the next section.

In this paper, we specifically explore the various types of excited states, namely, extended, localized and critically delocalized
states by studying the non-equilibrium dynamics of the model.
For this, we focus on the evolution of entanglement
entropy as a result of sudden quenches of different strengths to various phases of the Hamiltonian
 to explore the interplay of localization and delocalization in the ground and excited states.
The system is prepared in the ground state of the initial Hamiltonian $H_0$. At $t=0$, one of the parameters of the Hamiltonian is changed abruptly 
resulting to a new Hamiltonian $H$. The initial ground state is no longer the ground state of the final Hamiltonian
and the state of the system will dynamically evolve with respect to the Hamiltonian $H$. 
The non-equilibrium dynamics is 
studied by calculating the
evolution of the entanglement entropy of a subsystem of first $l-$spins
of the chain with the rest of the chain. The entanglement entropy $S_l(t)$ of $l-$spins forming the subsystem is defined as: 
$S_l(t)=Tr_l[\rho_l(t)\ln \rho_l(t)]$, i.e., tracing over all the sites greater than $l$, with $\rho_l(t)$being the reduced density matrix of the subsystem at time $t$ given by
$\rho_l(t)=Tr_{n\ne l} |\psi(t)\rangle \langle \psi(t)|$. Here $|\psi(t) \rangle$ is the state of the total system obtained by the evolution of the initial
ground state with respect to the new Hamiltonian $H$. It has been shown that in a homogeneous system which has extended states, $S_l(t) \propto t$
for $t < l/v_{m}$, where $v_m$ is the maximum velocity of the quasiparticles \cite{calabrese05}. For $t> l/v_{max}$, it saturates to an l-dependent
value. In case of random systems, $S_l(t)$ saturates almost immediately to a finite value due to the localized nature of the states
in disordered systems, whereas the behavior is ultraslow at the critical point \cite{igloi12}. 
Similar quench studies have also been performed on quasicrystals where $S_l(t) \sim t^{\sigma}$, with $0<\sigma<1$ \cite{igloi13}.
This paper aims to understand the complicated interplay of extended as well as localized states, i.e., the existence of the mobility edge
and its effect on the non-equilibrium dynamics of the model. To the best of our knowledge, this kind of study has not been done
in this model.
The techniques used for numerical calculations involve free fermions\cite{lieb61,igloi09}. We shall compare the evolution of $S_l(t)$ in QPTIM
with the known results in different types of phases, as discussed above. 

The paper is divided into the following sections. Section I consists of Introduction to the model,
with Section II describing the properties of Quasiperiodic transverse Ising model.
The results of the quench dynamics is presented in Section III after which we conclude the chapter with
the conclusions.

\section{The model}

\begin{figure}[h]
\begin{center}
\includegraphics[width=8.5cm]{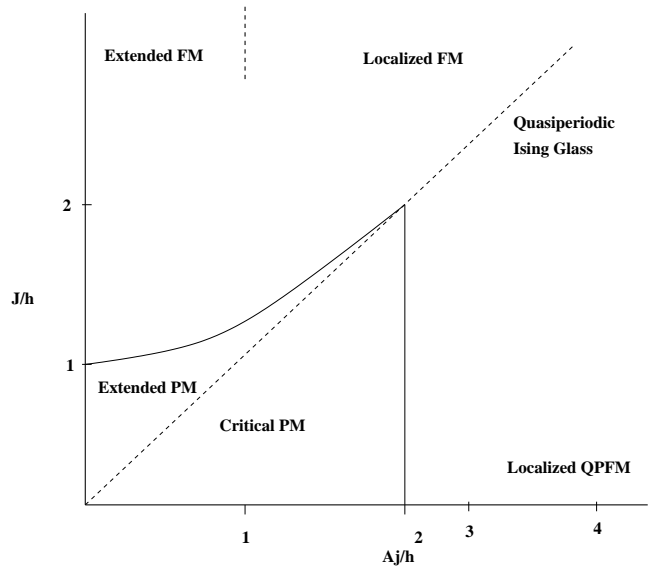}
\end{center}
\vskip -.5cm
\caption{Phase diagram of QPTIM consisting of FM, PM and QPFM ground states. Depending upon the strength of the quasiperiodic
modulation, the low energy excitations can exhibit localized, extended or critically delocalized behavior, also shown in the figure.
For more details, see Ref. \onlinecite{anushya17}.}
\label{fig_phasediagram}
\end{figure}

The Hamiltonian of QPTIM is given by
\begin{eqnarray}
H&=&-\frac{1}{2}\sum_j J_{j}\sigma_j^x \sigma_{j+1}^x + h \sigma_j^z,\nonumber\\
J_{j}&=&J+A_J\cos(Q(j+1/2))
\label{eq_ham1}
\end{eqnarray}
Here, $\sigma_j^\alpha$ are the Pauli matrices at site $j$, with $\alpha$
taking values $x,y,$ and $z$. For introducing quasiperiodicity in numerics,
we set the wave vector $Q$ to an incommensurate value given by
$Q=2\pi(\sqrt{5}+1)/2$, the golden ratio. It is to be noted that one can reach
the quasiperiodic limit by setting $Q=2\pi p/q$, with $p$ and $q$ as two consecutive numbers of
the Fibonacci sequence \cite{modungo09}. Unlike AA model which possess self duality, no such duality exists in
QPTIM, resulting to a much more rich phase diagram. The zero temperature phase diagram 
of this model as obtained by Anushya $et.al$ 
is presented in Fig.\ref{fig_phasediagram} which consists of three phases: paramagnetic (PM), ferromagnetic (FM) and quasiperiodically
alternating ferromagnet (QPFM). Depending upon the values of $J$ and $A_J$,
the excited states of the model can show extended, localized or critically delocalized
behavior. In the critically delocalized phase, the states have multifractal scaling behavior and hence this special
name.
The thick line originating from $J/h=1$ corresponds to a phase transition belonging to
the Ising universality class with $\nu=1$ and $z=1$. On the other hand, the second phase boundary separating critical PM
and Localized QPFM belongs to a different universality class with the same correlation length exponent as Ising critical point 
$i.e.,$ $\nu=1$, but with the dynamical exponent $z$ equal to $2$.
The ground state phase diagram and the properties of the excited states of this model 
have been obtained analytically only under certain limits where there exist enhanced symmetry.
This enables one to perform analytical calculations and comment on the energy independent 
features of the states.
Such special limits include $A_J=0$, $J=0$ and $J \to \infty$, and are briefly discussed below. 

{\bf{A brief discussion on the phase diagram}}

As mentioned before, we can analytically comment upon the localization-delocalization
properties of the eigenstates of the system only under certain limits,
the rest being an extrapolation of these analytical studies supported by numerics.
We start with the point $A_J=0$ and $J/h=1$ which corresponds to the well known Transverse Ising model 
critical point. It has gapless extended excitations at all energies. It is argued in Ref.\onlinecite{anushya17}
that the parabolic phase boundary originating from $J/h=1$, separating Ferromagnetic (FM) and Paramagnetic (PM)
phase, belongs to the same universality class as that of $A_J=0$ transverse field Ising model.
Therefore, atleast the low lying excitations along this phase boundary should be extended.
Also, since this phase boundary ends at $A_J=2$ and $J=2$, it need not be so 
beyond this terminating point. In the other extreme limit of $J \to \infty$, the ground state
of the model consists of all spins pointing along $+x$ (or $-x$) direction, which allows one
to re-write the Hamiltonian in terms of domain wall dynamics resembling AA model. Extending the AA model results to this point, we get the result that
all the states for $A_J<h$ are extended, and are localized for $A_J>h$. This limit corresponds to the dashed vertical line
shown in Fig. \ref{fig_phasediagram}. At the other extreme limit of $J =0$, there exist a triality at $A_J/h=2$ similar to 
the AA duality. For $A_J/h<2$, the states are critically delocalized whereas the spectrum is localized when $A_J/h>2$.
Energy independent localization properties are also present in the limit $J,A_J \ll h$ where extended PM to critically
delocalized PM transition occurs. All states are localized for $A_J \gg J,h$ \cite{anushya17}.
Other than these special points, the localization properties are generally claimed to be
$Q$ and energy dependent, which can be cross-checked through numerics. For more details, please refer to Ref. \onlinecite{anushya17}.

\begin{figure}[h]
\begin{center}
\includegraphics[width=9.2cm]{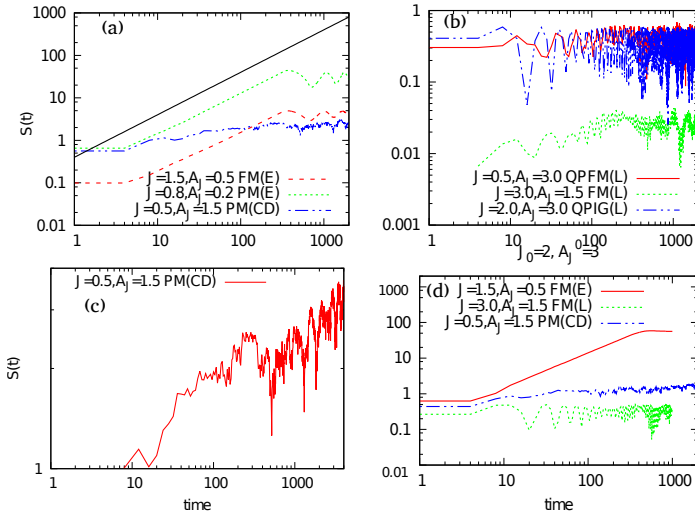}
\end{center}
\vskip -.5cm
\caption{Evolution of entanglement entropy in a log-log scale after quenching to different regions in the 
phase diagram starting from the same initial point $J_0=2$ and $A_J^0=0$ for a system of size $N=512$. 
The black continuous line
in Fig (a) corresponds to $S_l(t) \sim t$ confirming our claim of linear increase in $t$ when quenched to
the extended regime. Fig. (b) shows quenches to different localized phases. 
Fig. (c) highlights the slow increase of $S_l(t) \sim t^{\sigma}$ when quenched to a critically delocalized phase.
Fig. (d) corresponds to a different initial Hamiltonian $H_0$ (with $J_0=2,~A_J^0=3$) but showing behavior similar to Fig. (a) and (b)
thus confirming the fact that the nature of the evolution depends only upon the 
final Hamiltonian.
}
\label{fig_quench_2_0}
\end{figure}

\section{Sudden Quenches}
As discussed in the Introduction, we prepare the system in its ground state corresponding to a given $J_0$ and $A_J^0$, with $h$ set to unity.
At $t=0$, these parameters are instantaneously changed to $J$ and $A_J$, taking the system to some other point in the phase diagram.
The state of the system will now evolve  following the Schr\"odinger equation with the final Hamiltonian $H$.
In this paper, we study the evolution of entanglement entropy of a subsystem of the total system after such a quench. 

\begin{figure}[h]
\begin{center}
\includegraphics[width=6cm,angle=-90]{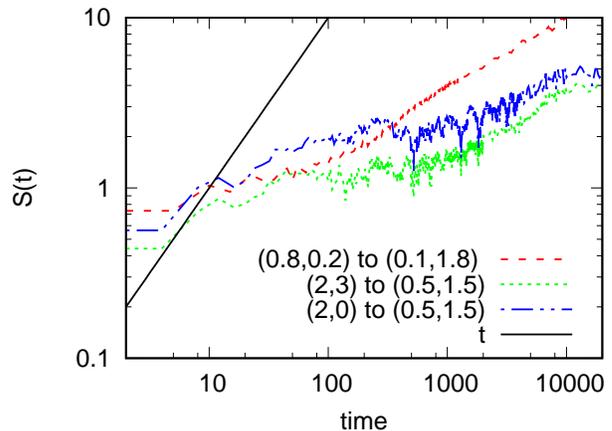}
\end{center}
\vskip -.5cm
\caption{Quenches of different strengths to critically delocalized phase. The parameters of the initial and final Hamiltonian are shown in the label.
For comparison, we have also plotted the black solid line
to depict the general linear increase in time when quenched to extended phase. 
Clearly, the increase in entanglement entropy seems to
be with an exponent smaller than unity. 
}
\label{critical}
\end{figure}

Since the phase diagram consists of localized, extended as well as critically delocalized excited states, we expect that the evolution of 
entanglement entropy will also capture the properties of these states. As mentioned before, entanglement entropy $S_l(t)$ of first $l$ spins
is $S_l(t)=Tr_{n \ne l}\rho_l(t) \ln \rho_l$. We have fixed $l=N/2$, where $N$ is the total number of spins.
The evolution of $S_{N/2}(t)$ is shown in Fig. \ref{fig_quench_2_0} for a given initial point $J_0=2$ and $A_J^0=0$
and different final points in the phase diagram. 
Depending upon the final Hamiltonian, we do observe that 
(i)When the final Hamiltonian has extended states, the $S_l(t)$ increases linearly with time as shown in Fig. \ref{fig_quench_2_0}a. Two
of the three quenches shown here are to the extended part of the phase diagram and they show the linear increase of
entanglement entropy with time. Similar behavior is also observed in quenches to other extended 
phase of the phase diagram independent of the initial point.
(ii) When the final Hamiltonian has localized states,
$S_l(t)$ saturates to a finite value almost immediately after the quench. 
This is presented in Fig. \ref{fig_quench_2_0}b. We find this behavior also to be
independent of the initial state of the system as expected. 
Fig. \ref{fig_quench_2_0}d shows the entanglement evolution for a different initial $H$
which has localized states after quenching to various points in the phase diagram, showing the same 
behavior as Fig \ref{fig_quench_2_0}a and b.
(iii) When the system is quenched to a phase with critically delocalized states,
there is an ultraslow increase of entanglement entropy with $S_l(t)\sim t^{\sigma}$ with $\sigma < 1$.
Unlike the quench to the extended states, the value of $\sigma$ when quenched to critically delocalized excited state region
seems to depend upon the parameters of the quench. We explored this non-universal $\sigma$ dependence further and the results
are shown in Fig. \ref{critical}. As shown, the increase of entanglement entropy is very slow as compared to quenches to the
extended phase. Also, one can clearly see from this figure that $\sigma$ is smaller than unity, the value of which varies from 
quench to quench.
We do want to emphasize here that all the numerical calculations
shown here are for $N=512$. Due to very slow increase of entanglement entropy when quenched to the critically 
delocalized region, the finite size effects may play an important role here.

\begin{figure}[tb]
\begin{center}
\includegraphics[width=5cm,angle=-90]{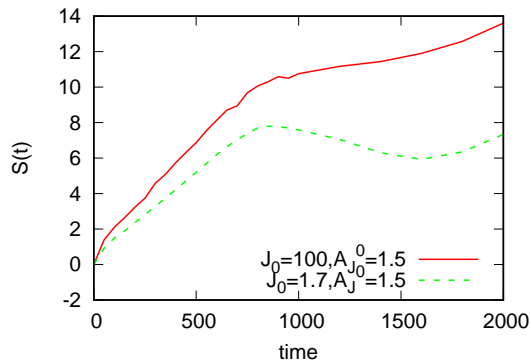}
\end{center}
\vskip -.5cm
\caption{Quench to $J=1.55$ and $A_J=1.5$ from two different initial points. The final point has extended low lying excited states
but have localized high energy excited states. A strong quench from $J_0=100, A_J^0=1.5$ is expected to explore the high energy localized states
which, in principle should be reflected in the evolution of entanglement entropy.
On the contrary, we see that such a strong quench is still
resulting to a linear increase in $S_l(t)$, similar to quench from $J_0=1.7,~A_J^0=1.5$ referred to as a weak quench.
}
\label{mobility}
\end{figure}

What will be interesting now is to check the possibility of 
a non-equilibrium evolution which explores both low-lying extended states and high energy localized states 
occurring at the same point
in the phase diagram, but controlled by the strength of the quench.
As per one of the figures in Ref. \onlinecite{anushya17}, the low lying states when $A_J=1.5$ and $1.5<J<1.7$ are extended
whereas the high energy states are localized. 
To explore this situation, we compare quenches to $J=1.55$ and $A_J=1.5$ which is extended paramagnetic phase, from two 
different initial points in the phase diagram. One of the quenches being a strong quench 
(since the parameters of the initial Hamiltonian is far away from that of final Hamiltonian) which is expected to explore high energies
whereas the other is a relatively weak quench which might explore only the low lying excited states.
The results of these two quenches are given in Fig. \ref{mobility}.
Both the quenches, strong and weak, are increasing almost linearly with time, contradicting the expectation of some effect of high energy localized
states. 
It seems to have no noticeable effect on the quench dynamics. 

\section{Conclusions}
We have presented the studies on non-equilibrium dynamics of quasiperiodic transverse Ising model
which shows localization protected excited state order without disorder. We have focussed on the 
nature of excited states and its effect on the non-equilibrium evolution of the initial state after a sudden quench,
where the initial state is chosen to be the ground state of the initial Hamiltonian. The important results from this study are
(i) Quench to regions with extended excited states shows an entanglement entropy which increases linearly
with time. (ii) Quench to regions with localized excited states shows an almost immediate 
saturation of entanglement entropy. (iii) Quench to critically delocalized state shows an increase of the
form $t^\sigma$, with $\sigma <1$, and depends upon the strength of the quench.
(iv) No noticeable effect is seen in the quench dynamics when the final Hamiltonian has both, extended low energy states and localized high
energy states. In both the cases, an almost linear increase of entanglement entropy is observed.
In future, we would like to explore the slow quenching dynamics of the same  model which is expected to show interesting results
atleast while quenching from localized to QPFM phase through a quantum critical point 
\cite{Sachdev,chakrabarti96,suzuki13} where the dynamical exponent is $z=2$.
This exponent plays an important role in the quenching dynamics
where a parameter of the Hamiltonian is varied slowly \cite{PolkovnikovRev,dziarmaga10,dutta15,mukherjee2007}.

\begin{acknowledgments}
UD acknowledges funding from DST-INSPIRE Fellowship by Government of India for financial support.
\end{acknowledgments}

\vspace{-\baselineskip}

\end{document}